\documentclass{article}
\usepackage{arxiv}
\usepackage[utf8]{inputenc} %
\usepackage[T1]{fontenc}    %
\usepackage{hyperref}       %
\usepackage{url}            %
\usepackage{booktabs}       %
\usepackage{amsfonts}       %
\usepackage{nicefrac}       %
\usepackage{microtype}      %
\usepackage{lipsum}
\usepackage{graphicx}
\usepackage{amsfonts} 
\usepackage{amsmath}
\usepackage{multicol}
\usepackage{multirow}
\usepackage{xcolor}
\usepackage{subcaption}
\usepackage{bbm}
\usepackage{enumitem}
\usepackage{bibentry}
\usepackage[switch]{lineno}
\usepackage{algorithm}
\usepackage{algorithmic}
\usepackage{cite}
\hypersetup{hypertex=true,colorlinks=true,urlcolor=black,linkcolor=blue,anchorcolor=blue,citecolor=blue}

\title{Click-through Rate Prediction with \\Auto-Quantized Contrastive Learning}
\author{
    Yujie Pan\thanks{Work done when Pan was an intern at Alibaba Group.} \\Shanghai Jiao Tong University\\\url{yujiepan@sjtu.edu.cn}
    \And
    Jiangchao Yao\\ DAMO Academy, Alibaba Group\\\url{jiangchao.yjc@alibaba-inc.com}
    \And
    Bo Han\\ Hong Kong Baptist University\\ \url{bhanml@comp.hkbu.edu.hk}
    \And
    Kunyang Jia \\ DAMO Academy, Alibaba Group \\  \url{kunyang.jky@alibaba-inc.com}
    \And
    Ya Zhang \\ Shanghai Jiao Tong University \\ \url{ya_zhang@sjtu.edu.cn}
    \And
    Hongxia Yang \\ DAMO Academy, Alibaba Group \\ \url{yang.yhx@alibaba-inc.com}
}

\begin{document}
\maketitle

\begin{abstract}
    Click-through rate (CTR) prediction becomes indispensable in ubiquitous web recommendation applications. Nevertheless, the current methods are struggling under the cold-start scenarios where the user interactions are extremely sparse. We consider this problem as an automatic identification about whether the user behaviors are rich enough to capture the interests for prediction, and propose an Auto-Quantized Contrastive Learning (AQCL) loss to regularize the model. Different from previous methods, AQCL explores both the instance-instance and the instance-cluster similarity to robustify the latent representation, and automatically reduces the information loss to the active users due to the quantization.  The proposed framework is agnostic to different model architectures and can be trained in an end-to-end fashion. Extensive results show that it consistently improves the current state-of-the-art CTR models.
\end{abstract}

\setcounter{footnote}{0}
\section{Introduction}
Click-Through Rate (CTR) prediction in the recommender systems is indispensable, which helps rank the candidate items meticulously based on the user interests. In the past few years, several Deep-Learning-based methods \textit{e.g.,} Wide\&Deep~\cite{cheng2016wide}, DeepFM~\cite{guo2017deepfm} and DIN~\cite{zhou2018deep}, have achieved impressive performance on this task. Nevertheless, these CTR models still suffer from the cold-start problem that breaks the assumption about the sufficient training samples available. Under the restricted user interactions in the cold-start scenarios, the model performance is dramatically limited~\cite{he2014practical}. This motivates a range of subsequent works~\cite{volkovs2017dropoutnet, lee2019melu, lu2020meta, zhu2021learning} to consider the cold-start recommendation.

One line of methods leverage the implicit regularization to prevent the CTR models from over-fitting~\cite{chen2019lambdaopt}. The popular techniques \textit{e.g.,} Dropout~\cite{srivastava2014dropout} and Early-stop~\cite{raskutti2014early}, would be considered during the training. For example, DropoutNet~\cite{volkovs2017dropoutnet} randomly disturbs the embedding of users or items to robustify the optimization procedure. Some other works explore to use the parameter or the embedding initialization to regularize the training of the CTR models~\cite{hospedales2021meta}. For example, MeLU~\cite{lee2019melu} learns to initialize the whole parameters of the models using MAML~\cite{finn2017model}. MAMO~\cite{dong2020mamo} extends MELU with two groups of memories to enhance the personalized initialization.
MetaEmb~\cite{pan2019warm}, MWUF~\cite{zhu2021learning} and GME~\cite{ouyang2021learning} explore to use side information, \textit{e.g.,} the item attributes and user neighbors, to initialize the user and item id embedding. For the recommendation model itself, the supervision for each sample is not changed during  training.

Another line of works explicitly construct the auxiliary task to help the training of the CTR models. For example, DeepMCP~\cite{ouyang2019representation} uses the additional tasks that model the user-item and item-item relationship to improve the user embedding and the item embedding. DIEN~\cite{zhou2019deep} and DMR~\cite{lyu2020deep} encourage the historical state representation close to the next click item, to better capture the evolution of user interests.
SSL4Rec~\cite{yao2020self} explores to use the unsupervised learning task, SimCLR~\cite{chen2020simple}, to enhance the generalization performance of the representation in the models. Similarly, CLCRec~\cite{wei2021contrastive} implements this goal by maximizing the mutual dependencies between item content and the collaborative signals. Empirically, these methods essentially incorporate the prior knowledge on the feature representation to help the training of the CTR models, which have achieved the state-of-the-art performance.

\begin{figure*}[t]
    \centering
    \includegraphics[width=0.5\linewidth]{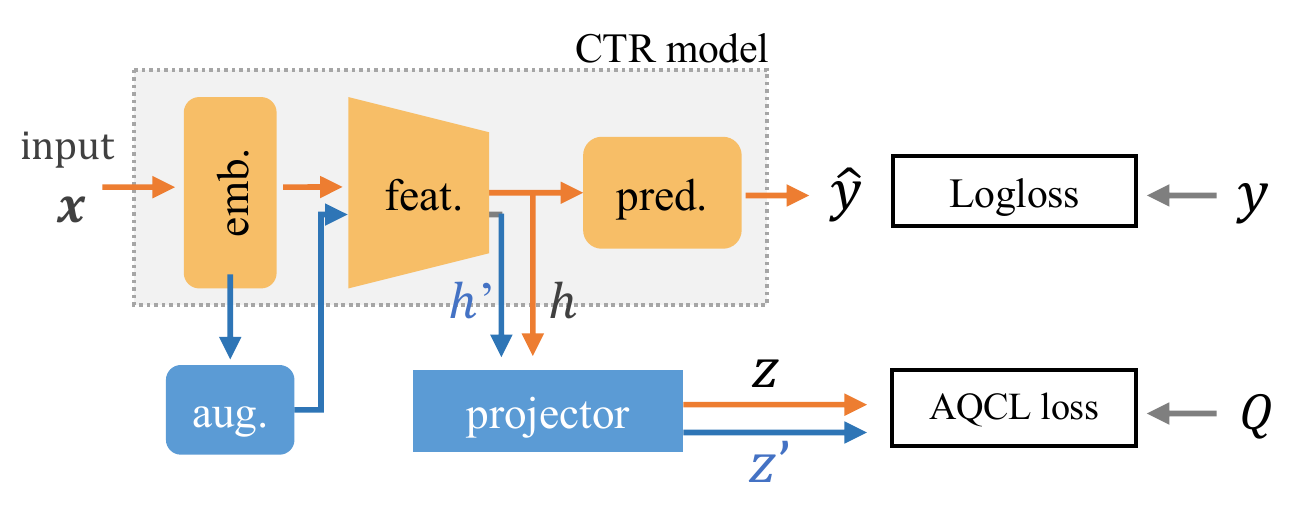}
    \hspace{0.01\linewidth}
    \includegraphics[width=0.4\linewidth]{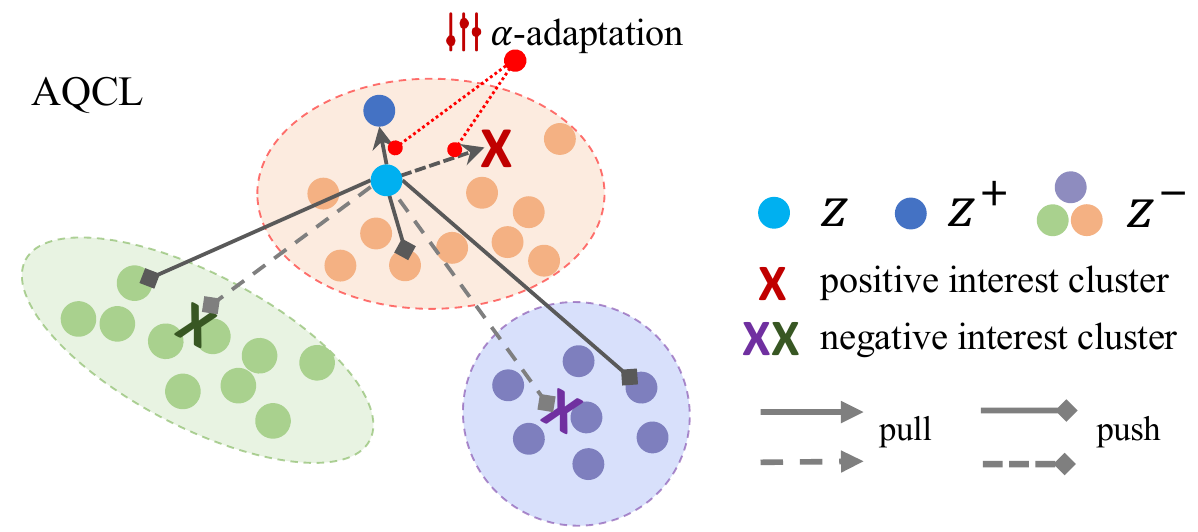}
    \caption{\textbf{Left}: Framework of Auto-quantized Contrastive Learning (AQCL) for the CTR model. For each input $x$, the embedding layer and feature interaction layer convert $x$ into latent code $h$. The primary  prediction task is guided by Logloss, using the output $\hat{y}$ and ground truth label $y$. Besides,
        a simple MLP projector $g(\cdot)$ outputs the representation $z = g(h)$. An augmented input is also fed similarly to get $z'$. The proposed AQCL loss uses the representations $z,\,z'$ and the interest cluster $Q$. During training, both Logloss and AQCL Loss are applied.
        \textbf{Right}: Motivation of AQCL. It achieves: (1) instance-instance similarity, where the representation $z$ should be close to $z^+$ from an augmented input and far from $z^-$ from negative sampling; and (2) interest clusters support, where $z$ should be close to the positive interest cluster and far from negative interest clusters among $Q$; (3) automatic balance of the instance-instance and the instance-cluster similarities for non-active/active users by $\alpha$-adaptation.}
    \label{fig:framework}
\end{figure*}

This work follows the second line and explores to leverage the structure of the representation space to automatically constrain the similarity among representations from different users. Specifically, we design an Auto-Quantized Contrastive Learning (AQCL) loss to regularize the training of the CTR models. Unlike the traditional contrastive learning approaches~\cite{chen2020simple, he2020momentum, chen2020big, chen2021empirical} and some attempts on the recommendation tasks~\cite{zhou2020s3, xie2020contrastive, yao2020self,wei2021contrastive} that focus only on instance-level discrimination, AQCL encourages both the instance-instance similarity and the instance-cluster similarity to automatically contribute to the modeling of the user interests. Figure~\ref{fig:framework} illustrates the framework and the intuition of AQCL.
We conduct a range of experiments on three sparse datasets and the results show AQCL consistently improves the CTR performance under the cold-start scenarios.

In total, our contributions can be summarized as follows:

\begin{itemize}[leftmargin=10pt]
    \item We introduce an auxiliary AQCL loss that automatically leverages the instance-instance similarity and the instance-cluster similarity to  regularize the representations in the CTR models under the cold-start scenarios.

    \item The interest clusters used in AQCL are learned together with the loss of the primary task in an end-to-end manner. Simultaneously, an $\alpha$-adaptation strategy is searched to automatically control the geometric balance for the representation of the active users and the non-active users.

    \item Extensive experiments on three datasets prove the effectiveness of our proposed method. Besides, AQCL is compatible with the common DNN-based CTR models like W$\&$D, DeepFM and DIN, which improves the ranking performance on both non-active users and active users.
\end{itemize}

\section{Related Works}

\subsection{CTR in Cold-Start Recommendations}
Recommender systems have been well studied in the past decades~\cite{deshpande2004item, he2014practical, rendle2010factorization, cheng2016wide, linden2003amazon, barkan2016item2vec, covington2016deep, yi2019sampling, zhang2021cause, yao2021device, kang2018self, tan2021sparse}, while the cold-start problem is a long-standing challenge in recommendation tasks. As mentioned before, many previous works can be considered as implicit or explicit regularization on the model.
The former tries to interfere with the optimization without extra tasks. For example, DropoutNet~\cite{volkovs2017dropoutnet} applies a data-augmentation on input to encourage robust user or item representations. Many meta-learning based works, \textit{e.g.}, MeLU~\cite{lee2019melu}, MataEmb~\cite{pan2019warm}, MAMO~\cite{dong2020mamo}, MetaHIN~\cite{lu2020meta}, PAML~\cite{wang2021preference},  GME~\cite{ouyang2021learning} and MWUF~\cite{zhu2021learning}, explore to initialize model parameters or embeddings with user and item side information. Some other works train the recommendation model with auxiliary task as explicit regularization.  DeepMCP~\cite{ouyang2019representation} is an early attempt to explore representation learning by designing matching subnet and correlation subnet.
SSL4Rec~\cite{yao2020self} and CLCRec~\cite{wei2021contrastive} applies traditional contrastive learning loss on either users or item representations. Our method also introduces an auxiliary task, but more comprehensively explores the representation space.

\subsection{Contrastive Learning}
Self-supervised Learning (SSL) is an unsupervised approach in learning data representations~\cite{liu2020self} and has shown success in computer vision~\cite{wang2019self, li2018non},  audio~\cite{baevski2019vq, ravanelli2020multi}, natural language processing~\cite{devlin2018bert, lan2020albert} and many cross-modality tasks~\cite{alwassel2020self, zhang2020devlbert, owens2018audio}.
Contrastive learning (CL) is one representative line of works, including CPC~\cite{oord2018representation}, MoCo~\cite{he2020momentum}, SimCLR~\cite{chen2020simple} and PIRL~\cite{misra2020self}. CL maximizes a lower bound on the mutual information between two or among more ``views" of an instance~\cite{wu2020on}. By identifying the positive sample pairs among other negative pairs, it succeeds in capturing the intrinsic features from individual instances in the latent space. Several attempts have used contrastive learning in sequential recommendations to learn either better item-level features~\cite{zhou2020s3,yao2020self,wei2021contrastive} or user representations~\cite{xie2020contrastive} individually, while our method considers the composed representations of the user and the item. Some works have also extended the traditional contrastive learning with more positive pairs. For example, SupCon\cite{khosla2020supervised} utilizes extra labels and make each instance close to others with the same class. PCL~\cite{li2020prototypical} uses EM to conduct unsupervised clustering and contrastive learning together. Similar to our AQCL method, these works also explore instance representations with other neighbors or clusters. However, it ignores the negative effect on the representation of active users who actually require the sufficient details in representation regarding the recommendation.

\section{Preliminary}
\subsection{CTR Prediction}
The CTR prediction as a binary classification problem is to find a map $f(\mathbf{x}_j)\rightarrow y_j$ for each pair $\left( \mathbf{x}_j, y_j  \right)\in \mathcal{D}$. Generally, for each input $\mathbf{x}_j$, it at least contains the user id $u_j$ and the candidate item id $i_j$.
Besides, the user historical clicks $\mathbf{s}_j = [i_{j,1}, i_{j,2},\cdots, i_{j,L_j}]$ are often considered, where $L_j$ is the length of the click sequence. Combining the user id $u_j$, the item id $i_j$, the historical clicks $\mathbf{s}_j$ and the other features $o_j$, we have $\mathbf{x}_j = (u_j, i_j, \mathbf{s}_j, o_j)$. The corresponding target label is a binary scalar $y_j \in \{0,1\}$ meaning whether the user $u_j$ clicks on the candidate item $i_j$.  A typical deep CTR model $f$ consists of the following parts \cite{zhang2021deep}:

\begin{itemize}[leftmargin=10pt]
    \item{\emph{Embedding layer.}} It transforms the sparse categorical features into dense-valued vectors \textit{i.e.,} embedding. Features like the item id are projected as the fixed-length embedding. For the historical sequence, we correspondingly acquire a sequence of embedding for the interacted items.

    \item{\emph{Feature interaction layer.}} The transformed embeddings are then fed into the interaction layers to produce a compact representation $\mathbf{h}_j$ for input instance $\mathbf{x}_j$. This component has diverse designs such as Multi-Layer Perception (MLP) \cite{guo2017deepfm}, Cross Network \cite{wang2017deep} and Multi-Head Self-Attention \cite{song2019autoint}.

    \item{\emph{Prediction layer}}. Finally, a simple prediction layer (usually a logistic regression module) %
    produces the final score $\hat{y}_j = f \left( \mathbf{x}_j \right)\in [0,1]$ for $\mathbf{x}_j$ on the representation $\mathbf{h}_j$.

\end{itemize}

With model output $f(\mathbf{x}_j)$ and ground truth $y_j$, the CTR model is trained on dataset $\mathcal{D}$ with the Logloss $\mathcal{L}_{\mbox{c}}$:
\begin{equation}
    \mathcal{L}_{\mbox{c}} = - \frac{1}{|\mathcal{D}|}
    \sum_{j=1}^{|\mathcal{D}|}
    \left(     y_j \log f\left( \mathbf{x}_j \right) +
    \left( 1 - y_j \right)  \log \left(1 - f(\mathbf{x}_j) \right)
    \right).
\end{equation}

\subsection{Cold-start Problem in CTR}

\begin{figure}[!h]
    \centering
    \includegraphics[width=0.25\linewidth]{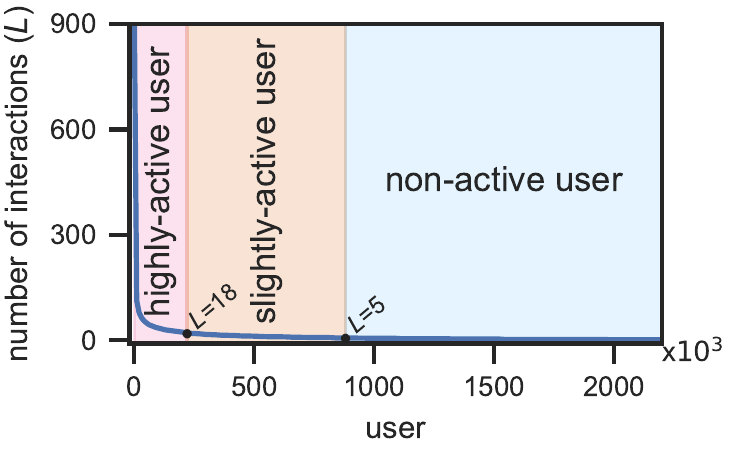}
    \caption{The user interactions (descending sort) on a cold-start industrial scenarios.}
    \label{fig:user-cdf}
\end{figure}

Users diverse a lot when it comes to the activeness in the cold-start scenarios\footnote{The cold-start scenarios here we mean are the early stage of the recommendation feed applications that have many slightly-active or non-active users but also have a small fraction of active users.}. We divide users into three groups, non-active, slightly-active and highly-active users based on the length of user click sequence $\mathbf{s}_j$. Figure~\ref{fig:user-cdf} plots the curve of the user sample number on one  cold-start industrial dataset, and categorizes three groups roughly of 60\%, 30\% and 10\% users. We can see that there are only a few of active users and a large proportion of users produces very limited interactions, making the CTR prediction task challenging.

\begin{figure}[tb]
    \centering
    \includegraphics[width=0.6\linewidth]{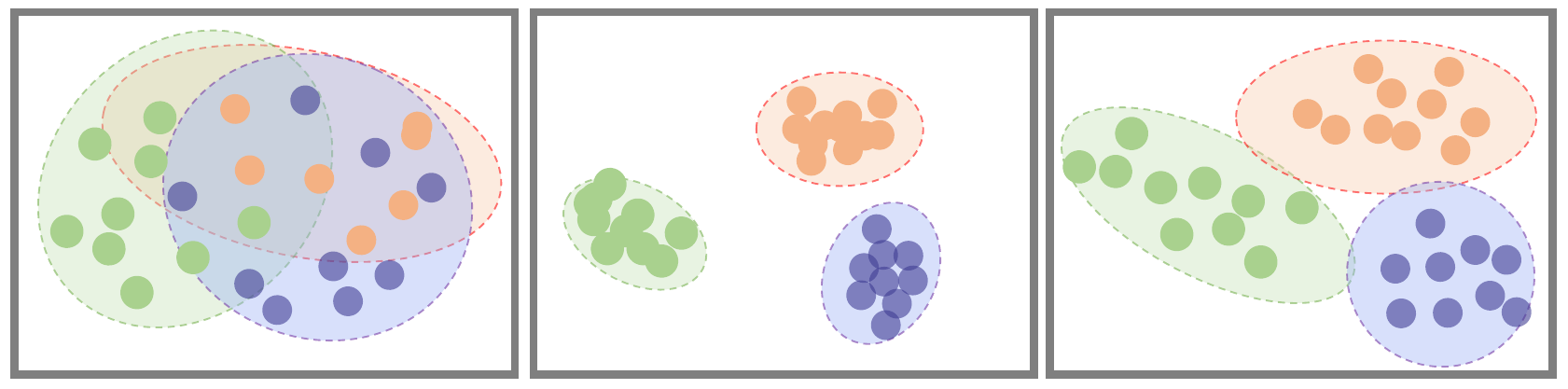}\\
    \begin{minipage}{0.2\linewidth}\centering (a) $\alpha=0$\end{minipage}
    \begin{minipage}{0.2\linewidth}\centering (b) $\alpha=1$\end{minipage}
    \begin{minipage}{0.2\linewidth}\centering (c) adaptive $\alpha$\end{minipage}
    \caption{Illustration of AQCL loss with different $\alpha$ in latent space. With $\alpha=0$, AQCL approximately degrades to ICL, which imposes each sample to be different from others. Ideally the representations are evenly distributed~\cite{wang2020understanding}. With $\alpha=1$, AQCL acts like quantized contrastive learning (QCL), which focuses on the interest clustering to help non-active users. The representations might lose the rich details for the CTR primary task. With an adaptive control of $\alpha$, AQCL combines the advantages of both ICL and QCL. It encourages the sample representations to  be  assigned to certain interest clusters, while the distances between samples are still guaranteed to maintain enough information for the CTR prediction.}
    \label{fig:comparison}
\end{figure}

\section{Auto-Quantized Contrastive Learning}\label{sqc}

\subsection{Self-supervision Framework for CTR\label{sec:ssl}}
In the cold-start scenarios, the click signal is usually scarce for training. In this case, the recent self-supervised learning (SSL), \textit{e.g.,} the contrastive learning loss SimCLR~\cite{chen2020simple}, is a straightforward choice as an auxiliary task to regularize the CTR model. In SimCLR, for each training instance $\mathbf{x}$ and its randomly augmented version $\mathbf{x}^+$, they go through the same feature interaction layer to get corresponding representations $\mathbf{z}$ and $\mathbf{z}^+$ after projector $g(\cdot)$. The classical loss focuses on the instance-level contrastive learning (ICL) to maximize the similarity between $\mathbf{z}$ and $\mathbf{z}^+$,
\begin{equation}\label{eq:icl}
    \mathcal{L}_{\textrm{ICL}}(\mathbf{z})=
    -\log \frac{\exp \left(\texttt{sim}\left(\mathbf{z}, \mathbf{z}_{}^{+}\right) / \tau\right)}
    {
        \sum_{\mathbf{z'} \in \{\mathbf{z}^+\} \cup \mathbf{Z}^-} \exp \left(\texttt{sim}\left(\mathbf{z}, \mathbf{z}' \right) / \tau\right)
    },
\end{equation}
where $\mathtt{sim}$ means cosine similarity, $\tau$ is a temperature hyper-parameter and $\mathbf{Z}^-$ are samples from negative sampling.
Regarding the CTR task, we applying the following input augmentation: first, we randomly mask some items from the user click history $\mathbf{s}_j$; and then we randomly set some embedding bits into zero, like dropout operation. Similar to \cite{chen2020simple}, we transform the latent code $\mathbf{h}$ by a small non-linear projection module $g(\cdot)$ to get the actual representation $\mathbf{z}$ for SSL task, i.e., $\mathbf{z}_i = g(\mathbf{h}_i)$. In practice, $g(\cdot)$ is implemented as a 3-layer MLP with the leaky-ReLU~\cite{maas2013rectifier} activation. The final training loss in the framework is a combination of both the primary CTR prediction task and the self-supervised auxiliary task. Note that, the SSL module only participates in the training stage.

\subsection{Auto-Quantized Contrastive Learning}
However, ICL loss only explores the instance-level similarity and fails to capture the relationship between neighbors. For non-active users, it is reasonable to get benefits from the neighbors with the rich behaviors in the latent space. This motivates us to model the structure information in the latent representation space, which can be also considered as the user interest clusters. Here, we define the codeword as interest in the representation codebook, borrowed from the concept of vector quantization~\cite{gray1984vector}, and propose an Auto-Quantized counterpart of contrastive learning.
Figure~\ref{fig:comparison} shows the difference between ICL and AQCL.
Formally, for $T$ interests $Q=[\mathbf{q}_1, \mathbf{q}_2, \cdots, \mathbf{q}_T]$ and a certain $\mathbf{z}$, we find the top-$K$ closest codewords $Q^+$, i.e.,
\begin{equation}
    Q^+ = {\arg\max} _{\mathbf{q}^1, \mathbf{q}^2,\cdots, \mathbf{q}^K \in {Q}} \sum_{k=1}^K \texttt{sim}(\mathbf{z}, \mathbf{q}^k). \nonumber
\end{equation}
Note that, \texttt{sim} is the cosine similarity in unit-sphere space as in Eqn.~\eqref{eq:icl}, since it is much less vulnerable to mode collapse~\cite{ma2019learning} and is widely used in prototype-based methods~\cite{li2020prototypical}. The proposed Auto-Quantized contrastive learning loss is formulated as a dynamic combination of instance-level and cluster-level contrastive learning:
\begin{equation}
    \label{AQCL}
    \begin{gathered}
        \mathcal{L}_{\textrm{AQCL}}  (\mathbf{z}) =
        -\log \frac{
            \left[ d_1\left(\mathbf{z}, \mathbf{z}^+ \right) \right] ^ {1-\alpha}
            \left[\sum_{\mathbf{q}^+ \in Q^+} d_2\left(\mathbf{z}, \mathbf{q}^+ \right) \right] ^ {\alpha}
        }
        {
            \sum_{ \mathbf{z'} \in  \{\mathbf{z}^+\} \cup {\mathbf{Z}^-}}  d_1 \left(\mathbf{z}, \mathbf{z}' \right)
            +
            \sum_{\mathbf{q'} \in Q}  d_2 \left(\mathbf{z}, \mathbf{q}' \right)
        }, \\
        d_1\left( \mathbf{z}, \mathbf{z}^{'} \right) = \exp \left(\texttt{sim}\left(\mathbf{z}, \mathbf{z}^{'}\right) / \tau_1 \right), \\
        d_2\left(\mathbf{z}, \mathbf{q}^{'} \right) = \exp \left(\texttt{sim}\left(\mathbf{z}, \mathbf{q}^{'}\right) / \tau_2 \right),
    \end{gathered}
\end{equation}

where $\tau_1$ and $\tau_2$ are temperature hyper-parameters for instances and clusters, and $\alpha$ controls the geometric mean of instance-instance and instance-cluster similarities. The design of $\alpha$ is for automatic balance for the representation personalization of non-active users and active users, which will be explained later. AQCL extends ICL by introducing positive support from both its augmented version $\mathbf{z}'$ and a set of $K$ closest codewords. We allow $K$ to be equal or larger than 1, because a sample representation may contain several interests, and using multiple codewords can be more stable to the probably incomplete interest clustering. Note that, the codewords are built based on the whole dataset, while the positive and negative pairs are from the same batch.

\subsubsection{Building the codebook $Q$}
In AQCL, a good codebook should try to cover all the sample representations with relatively small distances. Considering the possibly large-scale training dataset, we leverage an online method~\cite{Caron2020UnsupervisedLO} to learn the codebook along with AQCL training. In detail, for a codebook $Q=[\mathbf{q}_1, \mathbf{q}_2, \cdots, \mathbf{q}_T]$ and a batch of representation $Z = [\mathbf{z}_1, \mathbf{z}_2, \cdots, \mathbf{z}_B]$ with batch size $B$, we would acquire the corresponding assignment code matrix $A = [\mathbf{a}_1, \mathbf{a}_2, \cdots, \mathbf{a}_B] \in \mathbb{R}_+^{T\times B}$. Each column $\mathbf{a_b}$ of $A$ denotes the probability of assigning $\mathbf{z}_b$ into the totally $T$ codewords. Similar to \cite{asano2019self, Caron2020UnsupervisedLO}, the objective is
\begin{equation}
    \max_{A\in\mathcal{A}} \mbox{Tr}(A^\top Q^\top Z) + \epsilon H(A),
\end{equation}
where $H$ is the entropy function serving as a regularization with a small weight $\epsilon$. We define the constraints of $A$ by
\begin{equation}
    \mathcal{A} = \left\{ A \in \mathbb{R}_+^{T\times B} \mid
    A\mathbf{1}_B = \cfrac{1}{T} \mathbf{1}_T, Q^\top \mathbf{1}_T = \cfrac{1}{B}\mathbf{1}_B
    \right\}\,,  \nonumber
    \label{eqn:assign}
\end{equation}
where $\mathbf{1}_T$ denotes the vector of ones in dimension $T$. The constraint ensures that each codeword is roughly assigned evenly. This can be considered as an optimal transport problem \cite{asano2019self} and solved by the iterative Sinkhorn-Knopp \cite{cuturi2013sinkhorn} algorithm with the small computation cost. Then, we convert the continuous solution $A^*$ into its discrete, one-hot version using $\arg\max$ operation. For each representation $\mathbf{z}_b$, we encourage it to be close to only one of interest codewords. In summary, the loss to build the codebook is transformed as
\begin{align}\label{eqn:codebook}
    \begin{split}
        & \mathcal{L}_{\textrm{codebook}}  = -\sum_{b=1}^B \sum_{t=1}^T \mathbf{a}_b^{(t)} \log \mathbf{p}_b^{(t)}, \\
        & \mathbf{p}_b^{(l)} = \cfrac{  \exp (\texttt{sim} \left( \textrm{sg} (\mathbf{z}_b), \mathbf{q}_l \right)    / \tau_3 ) }{ \sum_{\mathbf{q}\in Q} \exp (\texttt{sim} \left( \textrm{sg} (\mathbf{z}_b), \mathbf{q} \right)    / \tau_3 )},
    \end{split}
\end{align}
where $\textrm{sg}(\cdot)$ means the stop-gradient operation, and $\tau_3$ is a temperature hyper-parameter. Note that, Eqn.~\eqref{eqn:codebook} is only to learn the codebook and not to update the model parameters.

\subsubsection{Auto-Quantization via $\alpha$-adaptation} Intuitively, users with the scarce clicks are more uncertain and need  support from their neighbors, and conversely, the performance of active users might suffer from over-quantization, since their representation maintains more details for prediction. Therefore, the representation learning should account for the different user activeness in the cold-start scenarios. To achieve this, we search $\alpha$ to automatically balance the importance of instance-instance measure and instance-cluster measure in Eqn.~\eqref{AQCL}.
Specially, we design a weight control module for $\alpha$ as a function of variable $L_j$ for user $u_j$, \textit{i.e.}, $\alpha_j = R(L_j)$. With $L_j$ increasing, we know better about the user history, and empirically we shall not enforce conformity and make $\alpha$ smaller. However, it is challenging and labor-consuming to design the weight \textit{vs.} activeness curve for each agnostic cold-start scenario. Therefore, we resort to AutoML~\cite{hutter2019automated} to search the appropriate function $R(L_j)$ for Eqn.~\eqref{AQCL} in the following.

\begin{figure}[h]
    \centering
    \includegraphics[width=0.23\linewidth, trim=0 5 0 5]{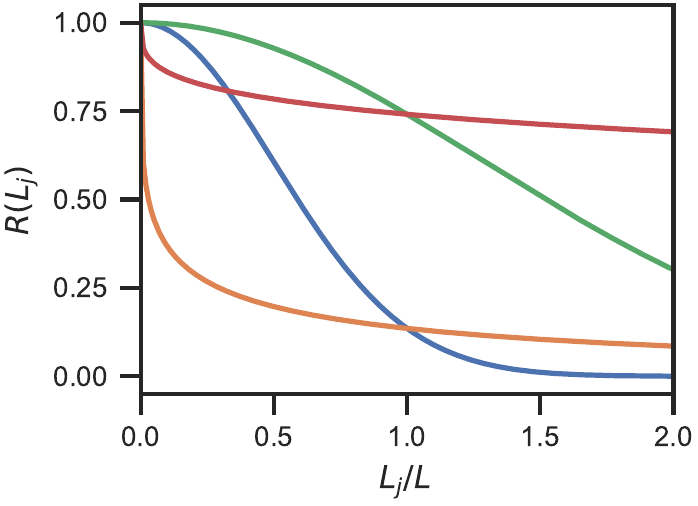}
    \caption{Possible search results for $R(L_j)$.}
    \label{fig:possible-search-results}
\end{figure}

\emph{Search space.}
First, the search space about $R(L_j)$ should take the following two intuitions into account: (1) when $L_j$ increases, $\alpha$ should be reduced; (2) $\alpha$ value should be in the range $[0,1]$. In this paper, we design the search space as
\begin{equation}\label{eq:search-space}
    \mathcal{F} =\left\{ R(L_j) = e^{-w_1 \cdot (L_j  / L)^{w_2}} : w_1>0, w_2 > 0\right\},
\end{equation}
where $L$ is the mean length of the click history for all users. The exact choice of basis function is not important. Figure~\ref{fig:possible-search-results} illustrates some possible search results.

\emph{Search objective.} For the problem in this section, we need a subset $\mathcal{D}_{\textrm{val}}$ (partitioned from the training set) to help auto-search. Given $\theta$ as the parameters of CTR model $f$, we target to search for the proper $\alpha$ function such that the model trained on training set $\mathcal{D}_{\textrm{train}}$ has the best performance on validation set $\mathcal{D}_{\textrm{val}}$.
Concretely, the objective is defined as
\begin{align}\label{eq:auto-ml}
    \begin{split}
        \{w_1^*, w_2^*\} & = \arg\min_{R(\cdot)\in \mathcal{F}}
        \mathcal{L}_{\textrm{val}} (f(\theta^*; R),  \mathcal{D}_{\textrm{val}}), \\
        & \theta^* = \arg\min_{\theta} \mathcal{L}_{\textrm{val}}(f(\theta; R), \mathcal{D}_{\textrm{val}}).
    \end{split}
\end{align}
where $\mathcal{L}_{\textrm{val}}$ is the Logloss $\mathcal{L}_c$ on validation set $\mathcal{D}_{\textrm{val}}$. Based on Eqn.~\eqref{eq:auto-ml}, we can find the optimal function in Eqn.~\eqref{eq:search-space}, which achieves the dynamic balance in representation learning.

\subsection{AQCL Algorithm}
The AQCL is implemented as an auxiliary loss to the primary CTR task. Therefore, the overall loss is
\begin{equation}
    \label{eqn:final-overall}
    \mathcal{L} = \mathcal{L}_{\textrm{c}} + w \mathcal{L}_{\textrm{AQCL}}\,,
\end{equation}
where $w$ is the weight for the auxiliary task. Note that, the codebook $Q$ is learned together with Eqn.~\eqref{eqn:final-overall} in an end-to-end manner, using the loss function in Eqn.~\eqref{eqn:codebook}. The training procedure is summarized in Algorithm~\ref{alg:algorithm}.
AutoML
eases the search of hyper-parameters $w_1, w_2$ in Algorithm~\ref{alg:algorithm} by using the objective~\eqref{eq:auto-ml}. Once the proper hyper-parameters are found by AutoML, we get the final CTR model. During the test phase, all components in the auxiliary task are omitted.

\begin{algorithm}[tb] {
        \caption{Algorithm for AQCL training}
        \label{alg:algorithm}
        \textbf{Input}: Training samples $\{\mathbf{x}_j\}$ and their history length $\{L_j\}$, \\
        \text{\quad\quad\,\,\,\,\,} parameters $w_1$ and $w_2$ for $\alpha$-adaptation\\
        \textbf{Output}: the CTR model
        \begin{algorithmic}[1] %
            \STATE Initiate CTR model $f$ and user interest codebook $Q$.
            \WHILE{not early stop}
            \STATE Fetch a batch of $\{\mathbf{x}_j\}$ and the click length $\{ L_j\}$.
            \STATE Get output $\{\hat{y}_j\}$ and projected representation $\{\mathbf{z}_j\}$.
            \STATE Get top-$K$ positive interests with codebook $Q$.
            \STATE Update $f$ by Eqn.~\eqref{eqn:final-overall} with $\alpha$ computed by Eqn.~\eqref{eq:search-space}.
            \STATE Get the discrete assignment matrix ${A}$ with Sinkhorn and update the user interest codebook $Q$ by Eqn.~\eqref{eqn:codebook}.
            \ENDWHILE
            \STATE \textbf{return} model $f$
        \end{algorithmic}}
\end{algorithm}

\section{Experiments}
In this section, we will evaluate the proposed AQCL framework. Specifically, we would like to answer the questions:

\begin{itemize}[leftmargin=10pt]
    \item{\textbf{RQ1.}} Compared with other methods, how does AQCL perform with the CTR model on different group of users?
    \item{\textbf{RQ2.}} Is AQCL as an auxiliary task generally compatible with the different CTR models?
    \item{\textbf{RQ3.}} How does AQCL work and how do the hyperparameters make the effect?

\end{itemize}

\subsection{Datasets}
We conduct experiments on three datasets with relatively severe data sparsity. The statistics are listed in Table \ref{dataset}.

\begin{table}[h]
    \centering
    \caption{Statistics for experiment datasets.}
    \renewcommand\arraystretch{1.05}
    \begin{tabular}{ccccc}
    \toprule
    dataset & \#users & \#items & \#samples &  \\
    \midrule
    Amazon  &        22,363    &  12,101     &    198,502        &  \\
    Ta Feng   &      32,266 &  23,812     &    817,741      &  \\
    Oncold  &       908,400  &  21,078      &    53,754,238     & \\
    \bottomrule
    \end{tabular}
    \label{dataset}
\end{table}

\begin{itemize}[leftmargin=10pt]
    \item{\textbf{Amazon}\footnote{\url{https://jmcauley.ucsd.edu/data/amazon}}.} This dataset~\cite{He2016UpsAD} is composed of product reviews from the Amazon website. We follow~\cite{chen2018neural,ijcai2018-521} and use the subset of Beauty to verify AQCL.  The task is defined to predict whether a user will comment about a certain item.

    \item{\textbf{Ta Feng}\footnote{\url{http://recsyswiki.com/wiki/Grocery_shopping_datasets}}.} This is a sparse grocery shopping dataset released by ACM RecSys. It covers products from food, office supplies to furniture. The dataset consists of user transactions from November 2000 to February 2001. We predict whether a user will buy a certain item.

    \item{\textbf{Oncold.}} This is an industrial dataset of the real-world online cold-start recommendation feeds, which is collected from May to July, 2021. The dataset is extremely sparse as most of users only have clicked a few of items.
\end{itemize}

Like~\cite{zhou2018deep, ma2019learning}, we sort the user behaviors by timestamp. For Amazon (Ta Feng, respectively), we use the last interaction (day) as the test, the second last interaction (day) as the validation, and the rest as the training data. For Oncold dataset, we split the data of last 20 days equally as the test set and the validation set, and the rest clicks are used as the training data.

\subsection{Experiment Settings}

\subsubsection{Baselines}
For RQ1, we compare with some representative methods of two research lines, \textit{i.e.,} the implicit regularization and the explicit regularization. For fair comparison, the following methods and AQCL all use DIN as the backbone model. Besides, we only refer to the design of regularization and omit their modifications of the CTR model architectures in these works, \textit{e.g.,} the positional encoding.
\begin{itemize}[leftmargin=10pt]
    \item \textbf{DropoutNet}~\cite{volkovs2017dropoutnet} is a training strategy that randomly masks user or item embeddings to handle the cold-start problem. It encourages the CTR model to make full use of the side information.
    \item \textbf{DeepMCP}~\cite{ouyang2019representation} is a representation-learning-aided model. Except the Logloss, it uses the matching subnet to capture the user-item relation, and the correlation subnet to explore the item-history relation.
    \item \textbf{DMR}~\cite{lyu2020deep} proposes an auxiliary matching loss to measure the correspondence between the user preference and the target item in the embedding space.
    \item \textbf{ICL}~\cite{chen2020simple} is the vanilla instance-level contrastive learning. We here use ICL during the training stage rather than pre-training.
\end{itemize}

For RQ2, we verify AQCL with the following backbones.
\begin{itemize}[leftmargin=10pt]
    \item \textbf{W$\&$D}~\cite{cheng2016wide} is a classical method that uses the feature-cross to help the model capture the high-order relationship hidden in the data for the better prediction.
    \item \textbf{DeepFM}~\cite{guo2017deepfm} is a successful attempt to combine the power of factorization machines in recommendation and deep learning in the feature learning.
    \item \textbf{DIN}~\cite{zhou2018deep} uses the attention mechanism to learn the representation of the user click history given the candidate item, to better explore the user interests.
\end{itemize}

\subsubsection{Implementations} All experiments are implemented in PyTorch and run on NVIDIA Tesla V100 GPUs.  We use Adam optimizer and the learning rate is $0.001$ during training. The dropout rate for DNNs is set as 0.2. We search the L2 regularization weight for the embedding among $\{10^{-5}, 10^{-4}, \cdots, 10^{-1}\}$ in all models. For AQCL, we define the weight $w$ of auxiliary task as $\{0.01, 0.05, 0.1\}$, and the temperatures $\tau_1, \tau_2, \tau_3$ are all set as 0.1. We set the codebook capacity $T$ as 128 for each dataset, and $K$ as 5.

\subsubsection{Evaluation metrics}
We adopt AUC and RelaImpr to evaluate the performance of the CTR models. AUC measures the probability that a randomly chosen positive sample is ranked higher than a randomly chosen negative sample. Higher AUC indicates better performance. RelaImpr, used in many works like~\cite{pmlr-v32-yan14, zhu2021learning}, shows the relative improvement of the target compared with the base. Here, we define the base as the backbone used in our experiments, \textit{e.g.,} DIN. RelaImpr is thus defined as
\begin{equation}
    \textrm{RelaImpr} = \left( \cfrac{\textrm{AUC (target)}-0.5}{\textrm{AUC (base)}-0.5} - 1\right) \times 100\%.
\end{equation}

We calculate the metrics on non-active, slightly-active and active users individually to monitor the model performance on the users with the different activeness.

\subsection{Results and Discussion}

\subsubsection{CTR prediction}
To answer RQ1, we conduct experiments on the baselines and our method. From Table \ref{result}, we can find AQCL achieves consistent improvements on three datasets.

\begin{table*}[t]
    \caption{The average result of 5 trials on Amazon, Ta Feng and Oncold datasets. DIN is the base without regularization.}
    \renewcommand\arraystretch{1.05}
    \centering
\begin{tabular}{cccccccccc}
\toprule
\multirow{2}{*}{Dataset} & \multirow{2}{*}{Model}   & \multicolumn{2}{c}{Overall}    & \multicolumn{2}{c}{Non-active user}           & \multicolumn{2}{c}{ Slightly-active user}  & \multicolumn{2}{c}{Highly-active user}   \\
&        & AUC     & RelaImpr& AUC     & RelaImpr& AUC     &  RelaImpr       & AUC     & RelaImpr\\
\midrule

\multirow{6}[0]{*}{Amazon} & DIN    & 0.6956 & 0.00\% & 0.6719 & 0.00\% & 0.7064 & 0.00\% & 0.7998 & 0.00\% \\
       & DropoutNet & 0.6929 & -1.38\% & 0.6691 & -1.63\% & 0.7034 & -1.45\% & 0.7975 & -0.77\% \\
       & DeepMCP & 0.7053 & 4.96\% & 0.6797 & 4.54\% & 0.7191 & 6.15\% & 0.8097 & 3.30\% \\
       & DMR    & 0.6982 & 1.33\% & 0.6736 & 0.99\% & 0.7087 & 1.11\% & 0.8064 & 2.20\% \\
       & ICL    & 0.7016 & 3.07\% & 0.6767  & 2.79\% & 0.7138  & 3.59\% & 0.8059  & 2.03\% \\
       & AQCL   & \textbf{0.7078} & \textbf{6.24\%} & \textbf{0.6826} & \textbf{6.22\%} & \textbf{0.7231} & \textbf{8.09\%} & \textbf{0.8105} & \textbf{3.57\%} \\

\midrule

\multirow{6}[0]{*}{Ta Feng} & DIN    & 0.6865  & 0.00\% & 0.6783  & 0.00\% & 0.6929  & 0.00\% & 0.7009  & 0.00\% \\
       & DropoutNet & 0.6812  & -2.84\% & 0.6762  & -1.18\% & 0.6843  & -4.46\% & 0.6927  & -4.08\% \\
       & DeepMCP & 0.6881  & 0.86\% & 0.6791  & 0.45\% & 0.6947  & 0.93\% & 0.7057  & 2.39\% \\
       & DMR & 0.6880  & 0.80\% & 0.6806  & 1.29\% & 0.6930  & 0.05\% & 0.7031  & 1.10\% \\
       & ICL    & 0.6914  & 2.63\% & 0.6812  & 1.63\% & 0.6931  & 0.10\% & 0.7032  & 1.14\% \\
       & AQCL   & \textbf{0.6931} & \textbf{3.54\%} & \textbf{0.6832} & \textbf{2.75\%} & \textbf{0.6969} & \textbf{2.07\%} & \textbf{0.7098} & \textbf{4.43\%} \\

\midrule

\multirow{6}[0]{*}{Oncold} & DIN    & 0.7601  & 0.00\% & 0.7361  & 0.00\% & 0.7608  & 0.00\% & 0.7787  & 0.00\% \\
       & DropoutNet & 0.7615  & 0.54\% & 0.7387  & 1.10\% & 0.7617  & 0.35\% & 0.7756  & -1.11\% \\
       & DeepMCP & 0.7675  & 2.85\% & \textbf{0.7409} & \textbf{2.03\%} & 0.7728  & 4.60\% & 0.7914  & 4.56\% \\
       & DMR & 0.7598  & -0.12\% & 0.7365  & 0.17\% & 0.7570  & -1.46\% & 0.7778  & -0.32\% \\
       & ICL    & 0.7640  & 1.50\% & 0.7367  & 0.25\% & 0.7706  & 3.76\% & 0.7895  & 3.88\% \\
       & AQCL   & \textbf{0.7691} & \textbf{3.46\%} & 0.7396  & 1.48\% & \textbf{0.7823} & \textbf{8.24\%} & \textbf{0.7968} & \textbf{6.49\%} \\

\bottomrule
\end{tabular}
\label{result}
\end{table*}

\begin{itemize}[leftmargin=10pt]
    \item In general, we observe the overall performance improvement with the implicit/explicit regularization. However, there is a slight performance drop for DropoutNet on Ta Feng and DMR on Oncold. Considering that DropoutNet emphasizes the importance of the side information, the performance may get hurt if there is no enough auxiliary information available. For DMR, the representation correspondence constraint between the user and the target item might
          be harmful to the active users. For other cases, the positive effects are shown on both non-active users and active users. This means that the proper regularization can help the CTR model in cold-start scenarios.

    \item  Our method outperforms the baselines in most cases. The advantage of AQCL is that it can automatically capture the interest clusters to support the non-active users and weaken the information loss for the representation of the active users via $\alpha$ adaptation. For Amazon dataset, we can see the significant improvement on non-active users, which shows the effectiveness by using neighbour information to alleviate the sparsity issue. For Ta Feng and Oncold datasets, there is more gain on the highly-active users. This can be attributed to that the proper relaxation to incorporate the instance-instance similarity in the AQCL loss yields a more robust representation.
\end{itemize}

\subsubsection{Different backbones}

To answer RQ2, we adapt AQCL with other backbones, \textit{i.e.,} W$\&$D and DeepFM to verify its effectiveness.  For both W$\&$D and DeepFM, we do not modify their wide or FM part, and apply AQCL to the output before the last linear layer on the deep side. Table \ref{tab:deepfm} summarizes the results of AQCL with different backbones. According to the table, AQCL consistently improves the backbones, which demonstrates its compatibility with the popular CTR models.

\begin{table}[t]
    \caption{Performance of AQCL with different backbones.}
    \renewcommand\arraystretch{1.0}
    \centering
    \begin{tabular}{cccc}
     \toprule
     Model & Amazon &  Ta Feng & Oncold  \\
     \midrule
     W$\&$D &  0.6803 &  0.6716   &  0.7547  \\
     AQCL$_{\textrm{W$\&$D}}$  & \textbf{0.6841} &  \textbf{0.6732} & \textbf{0.7578}  \\
    
     \midrule
     
     DeepFM  &  0.6810 & 0.6721 & 0.7633                              \\
     AQCL$_{\textrm{DeepFM}}$  &  \textbf{0.6848} & \textbf{0.6768} & \textbf{0.7659}  \\
    
     \midrule
     DIN & 0.6956 & 0.6865 & 0.7601 \\
     AQCL$_{\textrm{DIN}}$ & \textbf{0.7078} & \textbf{0.6931} & \textbf{0.7691}\\  
    
    \bottomrule
    \end{tabular}
    \label{tab:deepfm}
\end{table}

\begin{figure}[t]
    \centering
    \includegraphics[width=0.22\linewidth, trim=0 2 0 0]{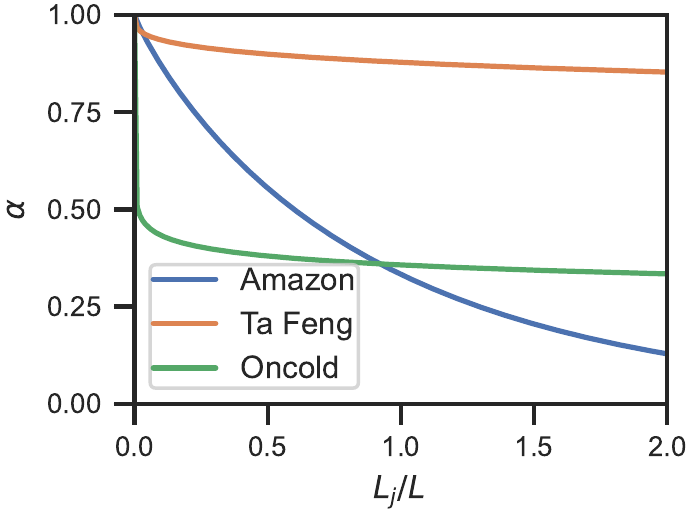}
    \hspace{0.01\linewidth}
    \includegraphics[width=0.22\linewidth, trim=0 2 0 0]{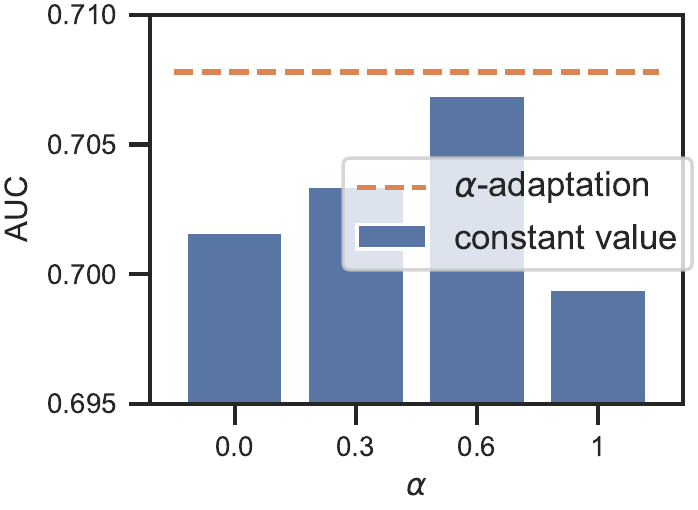}
    \hspace{0.03\linewidth}
    \caption{\textbf{Left}: The curves of $\alpha$ in AQCL for three datasets; \textbf{Right}: The experiments on Amazon with the constant $\alpha$.} \label{fig:automl}
\end{figure}

\subsection{Visualization and Ablation Study}
To answer RQ3, we conduct a range of visualization and ablation study about AQCL in the following.
\subsubsection{$\alpha$-adaptation}
The search result of $\alpha$ in AQCL for each dataset is plot in the left panel of Figure~\ref{fig:automl}. The curvatures are significantly different on three datasets. Amazon requires the relatively small instance-cluster regularization on active users. In comparison, Ta Feng emphasizes more dependency on the clusters and Oncold is similar but gives large $\alpha$ to all users.  Besides, we explore to replace $\alpha$-adaptation with the constant value in $\{0, 0.3, 0.6, 1.0\}$.  According to the right panel of Figure~\ref{fig:automl}, there is a performance drop compared with $\alpha$-adaptation, which demonstrates the effectiveness of AQCL to avoid the human labor.

\subsubsection{Representation $h$}
\begin{figure}[t]
    \centering
    \begin{subfigure}{0.22\linewidth}
        \includegraphics[width=\linewidth, trim=0 1 0 1]{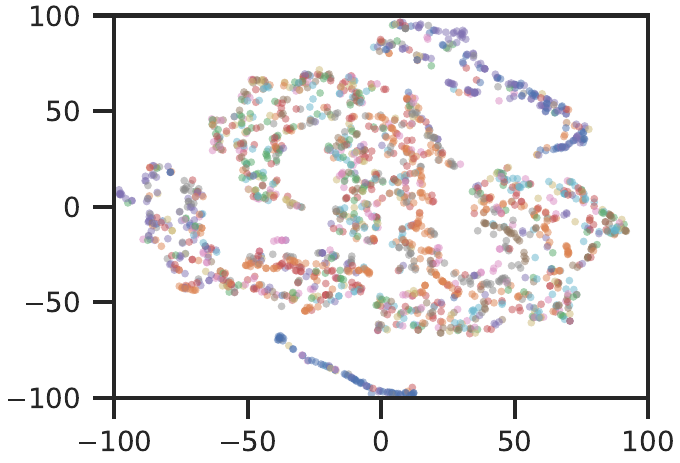}
        \caption{DIN} \label{fig:tsne-din}
    \end{subfigure}%
    \hspace{0.03\linewidth}
    \begin{subfigure}{0.22\linewidth}
        \includegraphics[width=\linewidth, trim=0 1 0 1]{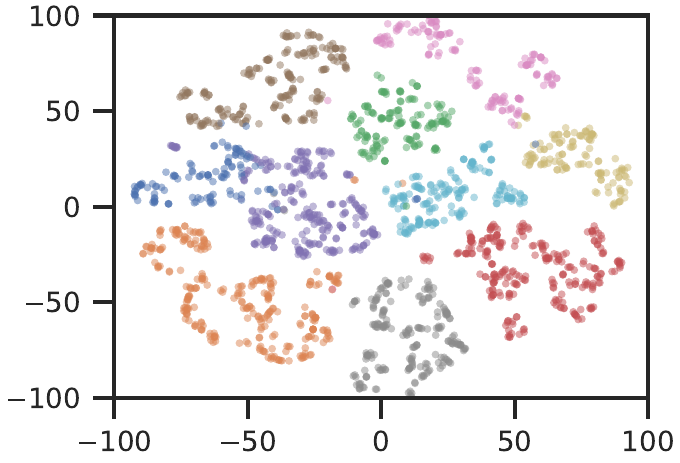}
        \caption{AQCL} \label{fig:tsne-aqcl}
    \end{subfigure}%
    \caption{t-SNE visualization of the latent representations respectively learned by DIN and AQCL on Oncold dataset.}\label{tsne}
\end{figure}

To visualize the learned representation in the latent space, we randomly choose a subset from Oncold dataset, and project the representations $\{\mathbf{h}_j\}$ into the 2D space via t-SNE~\cite{van2008visualizing} in Figure~\ref{tsne}. For a better view, we assign each point with the color representing the interest cluster deduced by AQCL. We find that AQCL can divide the sample representations roughly into several groups, while the vanilla DIN does not. This confirms the motivation of AQCL regarding the interest clusters, which might be useful to the non-active users.

\begin{figure}[t]
    \centering
    \begin{subfigure}{0.2\linewidth}
        \includegraphics[width=\linewidth, trim=0 1 0 1]{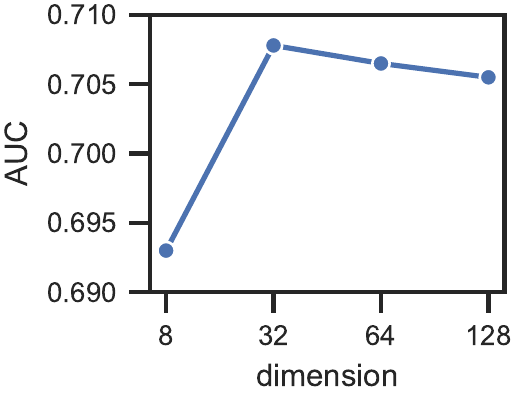}
        \caption{Dimension of $\mathbf{z}$} \label{fig:ablation-dimension}
    \end{subfigure}%
    \begin{subfigure}{0.2\linewidth}
        \includegraphics[width=\linewidth, trim=0 1 0 1]{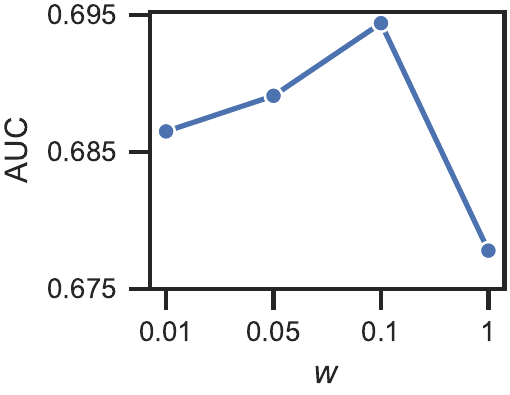}
        \caption{Weight of $\mathcal{L}_{\textrm{AQCL}}$} \label{fig:ablation-auxweight}
    \end{subfigure}%
    \begin{subfigure}{0.2\linewidth}
        \includegraphics[width=\linewidth, trim=0 1 0 1]{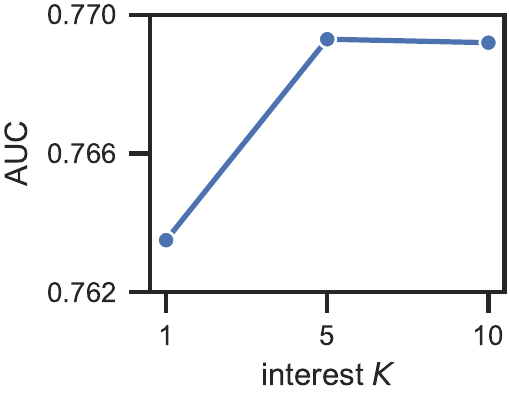}
        \caption{Top-$K$ interest} \label{fig:ablation-K}
    \end{subfigure}%
    \caption{(a) The effect of the dimension of $z$ on Amazon; (b) The effect of the auxiliary weight $w$ on Ta Feng. (c) The effect of the positive interest number $K$ on Oncold.} \label{fig:ablation}
\end{figure}

\subsubsection{Dimension of $z$}
As mentioned before, we project the hidden latent code $\mathbf{h}_i$ to another space by MLP $g(\cdot)$. The resulting vector $\mathbf{z}_i$ can have different dimensions. In Figure~\ref{fig:ablation-dimension}, we show the dimension of $\mathbf{z}$ is also important to the final performance. Specifically, we find that the dimension $\geq 32$ can keep the effectiveness of AQCL, while too small dimensions hinder the projector from collecting the enough information for the auxiliary task, and thus causes negative effects.

\subsubsection{Weight of $\mathcal{L}_{\textrm{AQCL}}$}
In this part, we conduct experiments with the different auxiliary task weight $w$ on Ta Feng dataset to verify its effect. As shown in Figure~\ref{fig:ablation-auxweight}, we empirically find that $w$ should not be very large. This might be because AQCL here serves as a data-driven regularization, and too large $w$ may result in the little attention to the primary task.
Besides, $w$ should not be too small, since in this case, it will degrade to the vanilla CTR task without the auxiliary gain.

\subsubsection{Interest number $K$} In AQCL, it allows each sample to be assigned into several interest clusters by adjusting $K$. Figure~\ref{fig:ablation-K} shows the effect of changing $K$ as $\{1,5,10\}$. We see that the performance decreased when $K=1$. This implies the user representation consists of multiple interests.

\section{Conclusion}
This paper aims at handling the cold-start scenarios to help the CTR model by designing an auxiliary task. We propose an Auto-quantized Contrastive Learning (AQCL) loss to encourage the model to leverage the possible interest clusters to help the non-active users and maintain the generalization ability to the active users. By training the CTR models with AQCL, we demonstrate our method consistently improve the current models, especially in the face of scarce interactions. The proposed framework is compatible with different model architectures and can be trained in an end-to-end fashion. We hope our work can inspire more explore to improve the CTR models with self-supervised representation learning.

\bibliographystyle{plain}
\bibliography{bib}
\end{document}